\documentclass[aps,pra,showpacs,twocolumn]{revtex4}
\usepackage{amssymb}
\usepackage{amsmath}
\usepackage{graphicx}
\usepackage{epsfig}

\begin{document}

\title{Relation between two measures of entanglement in spin-$1/2$ and
spinless fermion quantum chain systems}
\author{Xiao-Feng \surname{Qian}}
\author{Z. \surname{Song}}
\email{songtc@nankai.edu.cn}
\affiliation{Department of Physics, Nankai University, Tianjin 300071, China}

\begin{abstract}
The concepts of concurrence and mode concurrence are the measures of
entanglement for spin-$1/2$ and spinless fermion systems respectively. Based
on the Jordan-Wigner transformation, any spin-$1/2$ system is always
associated with a fermion system (called counterpart system). The comparison
of concurrence and mode concurrence can be made with the aid of the
Marshall's sign rule for the ground states of spin-$1/2$ $XXZ$ and spinless
fermion chain systems. We observe that there exists an inequality between
concurrence and mode concurrence for the ground states of the two
corresponding systems. The spin-$1/2$ $XY$ chain system and its spinless
fermion counterpart as a realistic example is discussed to demonstrate the
analytical results.
\end{abstract}

\maketitle

\section{introduction}

Quantum entanglement is one of the most intriguing features of quantum
mechanics for many body systems such as spin, fermion, boson systems and
etc. It plays a fundamental role in quantum information processing (QIP)
schemes and quantum computing, thus is regarded as an important resource for
these technologies \cite{chung}. It also has been demonstrated that the
quantum entanglement can be used to realize quantum teleportation \cite%
{bennett,Zeilinger,kimble,matini}, as well as to characterize the quantum
critical phenomena in strongly correlated systems \cite{Sachdev,
Osterloh:02,Wang, Gu, Momentum}.

In characterizing the pairwise entanglement in spin-$1/2$ systems, Wootters
proposed a definition of entanglement measure named concurrence \cite%
{Wootters:98}, which is easy to handle analytically and well accepted by the
quantum community. This definition, however, relies on the tensor product
structure of the state space with respect to a composite quantum system,
which, due to quantum statistics, does not appear obviously for systems of
indistinguishable particles, such as fermions and bosons. As a result, many
recent efforts are devoted to understand the quantum entanglement of
indistinguishable particle systems \cite{Lloyd, Burkard, Knill, Schliemann,
Li01, Paskauskas, Zanardi:01, Gittings, Eckert, Terhal, Zanardi:02, x wang,
Dunning, Zanardi, Barnum, Vedral, Shi03, Shi04}. Especially, a measure of
mode entanglement was proposed by Zanardi and Wang in grand canonical
ensembles based on the isomorphism between the full Fock space and qubit
space \cite{x wang}. It is recognized that the quantum entanglement is a
relative concept and its definition relies on the tensor-product structure
of the Fock space of indistinguishable particle systems. Namely, the same
Fock space, with respect to different single particle vector basis or the
normal modes, simply called modes, can be endowed with different tensor
product structures. With these recognitions, it\ is found that modes, rather
than particle labels, are universal concepts to properly describe the
quantum entanglement of indistinguishable particle systems. This idea is
used to characterize the ground-state entanglement of the BCS model \cite%
{Dunning} and superconductivity \cite{x wang,Shi04}, and also it is
investigated to derive the interesting relation between entanglement and
quantum state transfer \cite{qian}. Some related works were also proposed by
different authors \cite{Zanardi, Barnum, Vedral, Shi03} recently. From these
previous works, we notice that it is a common wisdom to treat mode
entanglement in the way of distinguishable particle, such as spin-$1/2$,
entanglements for the relativity of entanglement \cite{Zanardi:01}. However,
to our best knowledge, few papers are devoted to investigate the specific
relation between the entanglements of distinguishable particle and
indistinguishable particle systems. Therefore, in the present paper, we will
focus on the issue concerning the relation between spin-$1/2$ and spinless
fermion entanglements.

Although the two measures of entanglement are defined in different quantum
systems (spin-$1/2$ and spinless fermion systems), they are closely related
to each other based on the Jordan-Wigner transformation \cite{J-W}. With
this transformation, spin-$1/2$ operators can be represented by fermion
operators. Correspondingly, any spin model has a spinless fermion model as a
counterpart, i.e., if we span the Hilbert and Fock space with the basis $%
\{\left\vert m_{1}\text{, }...\text{, }m_{l}\text{, }...\text{, }%
m_{N}\right\rangle _{s}\}$, $m_{l}=\uparrow $, $\downarrow $ and $%
\{\left\vert n_{1}\text{, }...\text{, }n_{l}\text{, }...\text{, }%
n_{N}\right\rangle _{f}\}$, $n_{l}=1$, $0$ respectively, the two
Hamiltonians will share the same matrix form with each other in their
corresponding subspaces. Thus the two Hamiltonians have the common vectors
as eigen states corresponding to their own basis. In this way, the two
systems together with the Jordan-Wigner transformation have provided us a
perfect platform to investigate the differences as well as connections
between the two different definitions of measure of entanglement so as to
lead to some new insights in understanding the nature of spin and fermion
systems and also the nature of entanglement \cite{Zanardi:01,x wang}.

In this paper we will explore the relation between the two measures for
quantum chain systems. The key to these observations is the Jordan-Wigner
mapping of spins into lattice fermions. We will show that concurrence C \cite%
{Wootters:98} of a spin-$1/2${\Huge \ }$XXZ$\ chain state is different in
magnitude from the mode concurrence (MC) \cite{x wang} of the spinless
fermion counterpart state. This difference between the two measures is
directly related to, and thus to some extend,\ is a readout of, the
commutation relations of spins and fermions, which are the fundamental
features of the two distinct models. Our result is in agreement with the
statement in Ref. \cite{Shi03} that the entanglement is related to the
single-particle basis chosen. Furthermore, with the aid of Marshall's sign
rule \cite{Marshall, sign rule}, we reveal that there exists a simple
relation between the C and MC for the ground state of the spin-$1/2$ $XXZ$
chain model and its counterpart, spinless fermion model. The detailed
relations depend on different types of pairwise entanglement we concern. We
find that for the ground state of a spin-$1/2$ $XXZ$\ chain system and its
corresponding ground state of a many-particle spinless fermion system, (i)
the C between nearest neighbor (NN) sites in spin-$1/2$ $XXZ$ chain systems
is identical in magnitude to the MC in the corresponding spinless fermion
systems, (ii) C is no less than MC between any two non-NN sites. To
demonstrate our analytical results, some simple and realistic examples will
be discussed in detail at the end of the paper.

\section{General comparison of the two measures}

\subsection{Definitions of C and MC}

We first present the definitions of the concurrence and the mode concurrence
in spin-1/2 and spinless fermion systems respectively. Consider an $N$-site
spin-$1/2$ system
\begin{equation}
H_{s}=H_{s}(\{\sigma _{i}^{\alpha }\})  \label{Hs}
\end{equation}%
where $\sigma _{i}^{\alpha }$ is the Pauli matrix at $i$-th site$\ $and $%
\{\sigma _{i}^{\alpha }\}=\{\sigma _{i}^{\alpha }\mid \alpha =x,y,z;i\in
\lbrack 1,N]\}$. In this paper, we study the Hamiltonian that the $z$%
-component of the total spin is conserved, i.e., $[S^{z},H_{s}]=0$, where $%
S^{z}=\sum\nolimits_{i}S_{i}^{z}$ and $S_{i}^{\alpha }=\sigma _{i}^{\alpha
}/2$ for $\alpha =x,y,z$. Here and later in the paper we define that $\sigma
_{i}^{\pm }=\sigma _{i}^{x}\pm i\sigma _{i}^{y}$ and $S_{i}^{\pm
}=S_{i}^{x}\pm iS_{i}^{y}.$ Then the reduced density matrix of the eigen
state $\left\vert \psi _{s}\right\rangle $\ with respect to two arbitrary
sites\ $i$ and $j$ on the basis $\{\left\vert \uparrow \uparrow
\right\rangle $, $\left\vert \uparrow \downarrow \right\rangle $, $%
\left\vert \downarrow \uparrow \right\rangle $, $\left\vert \downarrow
\downarrow \right\rangle \}_{ij}$ is given by

\begin{eqnarray}
\rho _{s}^{ij} &=&tr_{N-2}(\left\vert \psi _{s}\right\rangle \left\langle
\psi _{s}\right\vert ) \\
&=&\left(
\begin{array}{cccc}
u^{+} &  &  &  \\
& \omega _{1} & z^{\ast } &  \\
& z & \omega _{^{2}} &  \\
&  &  & u^{-}%
\end{array}%
\right) ,  \notag
\end{eqnarray}%
where $\uparrow $ ($\downarrow $) denotes the spin up (down) and $tr_{N-2}$
means tracing over all the variables but the two on the sites $i$, $j$. Thus
the concurrence \cite{entangled rings} of the two separated sites $i$, $j$ is

\begin{equation}
C_{i,j}=2\max \left\{ 0,\left\vert z\right\vert -\sqrt{u^{+}u^{-}}\right\} ,
\end{equation}%
where, as given in Ref. \cite{Wang},

\begin{eqnarray}
u^{+} &=&\frac{1}{4}[1+(\left\langle \sigma _{i}^{z}\right\rangle
+\left\langle \sigma _{j}^{z}\right\rangle )+\left\langle \sigma
_{i}^{z}\sigma _{j}^{z}\right\rangle ]  \label{cfs} \\
u^{-} &=&\frac{1}{4}[1-(\left\langle \sigma _{i}^{z}\right\rangle
+\left\langle \sigma _{j}^{z}\right\rangle )+\left\langle \sigma
_{i}^{z}\sigma _{j}^{z}\right\rangle ]  \notag \\
z &=&\frac{1}{4}\left\langle \sigma _{i}^{+}\sigma _{j}^{-}\right\rangle ,
\notag
\end{eqnarray}%
are the expectations of Pauli matrices.

On the other hand, consider an $N$-site spinless fermion system with the
Hamiltonian

\begin{equation}
H_{f}=H_{f}(\{a_{i}^{\dagger }a_{j}\}),  \label{Hf}
\end{equation}%
where $a_{i}^{\dagger }$ is the fermion operator at the $i$-th site, and $%
\{a_{i}^{\dagger }a_{j}\}=\{a_{i}^{\dagger }a_{j}\mid i,j\in \lbrack 1,N]\}$%
. In this paper, we study the Hamiltonian that the total particle number $%
\hat{N}=\sum\nolimits_{i}\hat{n}_{i}=\sum\nolimits_{i}a_{i}^{\dagger }a_{i}$
is conserved, i.e., $[\hat{N},H_{f}]=0$. According to Ref. \cite{x wang},
the second order reduced density matrix of the state $\left\vert \psi
_{f}\right\rangle $\ with respect to two arbitrary sites\ $i$ and $j$ on the
basis $\{\left\vert 11\right\rangle $, $\left\vert 10\right\rangle $, $%
\left\vert 01\right\rangle $, $\left\vert 00\right\rangle \}_{ij}$ is given
as
\begin{eqnarray}
\rho _{f}^{ij} &=&tr_{N-2}(\left\vert \psi _{f}\right\rangle \left\langle
\psi _{f}\right\vert )  \label{elements} \\
&=&\left(
\begin{array}{cccc}
X^{+} &  &  &  \\
& Y^{+} & Z^{\ast } &  \\
& Z & Y^{-} &  \\
&  &  & X^{-}%
\end{array}%
\right) ,  \notag
\end{eqnarray}%
where $\left\vert \psi _{f}\right\rangle $ is an eigen state of the
Hamiltonian $H_{f}$ and similarly, $tr_{N-2}$ means tracing over all the
variables except the two on the sites $i$, $j$. Similarly, the nonzero
elements are determined by the correlation functions of the fermion operators

\begin{eqnarray}
X^{+} &=&\left\langle \hat{n}_{i}\hat{n}_{j}\right\rangle ,  \label{cff} \\
X^{-} &=&1-\left\langle \hat{n}_{i}\right\rangle -\left\langle \hat{n}%
_{j}\right\rangle +X^{+},  \notag \\
Z &=&\left\langle a_{i}^{\dagger }a_{j}\right\rangle .  \notag
\end{eqnarray}%
Correspondingly, the MC between sites $i$ and $j$ can be written as \cite{x
wang}

\begin{equation}
(MC)_{i,j}=2\max \left\{ 0,\left\vert Z\right\vert -\sqrt{X^{+}X^{-}}%
\right\} .  \label{concurrence}
\end{equation}

So far we have defined the measures of entanglement both in spin-1/2 and
spinless fermion systems respectively. In the following sub section we will
discuss the relations between the two measurements.

\subsection{Relations between C and MC for quantum chain systems}

As the two definitions of the concurrence correspond to different states and
models, we investigate the relation between C and MC based on two typical
models, spin-$1/2$ and spinless fermion chains. First, we concern an $N$%
-site $XXZ$ spin chain model, of which the Hamiltonian is given by

\begin{equation}
H_{s}^{XXZ}=%
\sum_{j=1}^{N-1}[J_{j}(S_{j}^{+}S_{j+1}^{-}+S_{j}^{-}S_{j+1}^{+})+J_{j}^{z}S_{j}^{z}S_{j+1}^{z}],
\label{xxz h}
\end{equation}%
where $J_{j}$ and $J_{j}^{z}$ are the coupling constants. Obviously, the $z$%
-component of the total spin is conserved for this model. The Hilbert space
of such a Hamiltonian is spanned by $2^{N}$ basis vectors $\{\left\vert
m\right\rangle _{s}\equiv $ $\left\vert
m_{1},...,m_{l},...,m_{N}\right\rangle _{s}\},$ where $m=1$, $2$, $3$, $...$%
, $2^{N}$, $m_{l}=\uparrow $, $\downarrow $ and $l=1$, $2$, $3$, $...$, $N$.
Then an arbitrary state of the above Hamiltonian can be generally written as

\begin{equation}
\left\vert \psi _{s}\right\rangle =\sum_{m}\gamma _{m}\left\vert
m\right\rangle _{s}\equiv \sum_{m}\gamma _{m}\otimes
_{l=1}^{N}(S_{l}^{+})^{f(m_{l})}\left\vert 0\right\rangle _{s},
\label{spin state}
\end{equation}%
where $\gamma _{m}$ are normalized coefficients, $\left\vert 0\right\rangle
_{s}\equiv \left\vert \downarrow \downarrow ...\downarrow \downarrow
\right\rangle _{s}$ represents the saturated ferromagnetic state, $S_{l}^{+}$
is the raising operator of the $l$-th qubit and the function $f(m_{l})=1,0$
correspond to $m_{l}=\uparrow $, $\downarrow $ respectively. We emphasize
that $\otimes _{l=1}^{N}(S_{l}^{+})^{f(m_{l})}$ is arranged in the ascending
order of $l$.

On the other hand, as is well known, there exists a counterpart spinless
fermion model for an arbitrary spin-1/2 $XXZ$ model. We now concern the
spinless fermion model obtained by employing the Jordan-Wigner
transformation \cite{J-W}%
\begin{eqnarray}
S_{l}^{+} &=&a_{l}^{\dagger }\exp (i\pi \sum_{p=1}^{l-1}\hat{n}_{p}),
\label{J-W} \\
S_{l}^{-} &=&\exp (-i\pi \sum_{p=1}^{l-1}\hat{n}_{p})a_{l},  \notag \\
S_{l}^{z} &=&\hat{n}_{l}-\frac{1}{2}.  \notag
\end{eqnarray}%
Based on the transformation, the corresponding spinless fermion Hamiltonian,
which is transformed from Eq. (\ref{xxz h}), is given as
\begin{eqnarray}
H_{f}^{TB} &=&\sum_{j=1}^{N-1}[J_{j}^{z}\left( \hat{n}_{j}-\frac{1}{2}%
\right) \left( \hat{n}_{j+1}-\frac{1}{2}\right)  \label{tb} \\
&&+J_{j}(a_{j}^{\dagger }a_{j+1}+\text{H.c.})].  \notag
\end{eqnarray}%
This is a typical tight-binding (TB) model with NN interaction $J_{j}^{z}$
and NN hopping integral $J_{j}$. The Fock space of the above Hamiltonian is
spanned by $2^{N}$ basis vectors $\{\left\vert n\right\rangle _{f}\equiv
\left\vert n_{1}\text{, }...\text{, }n_{l}\text{, }...\text{, }%
n_{N}\right\rangle _{f}\},$ where $n=1$, $2$, $3$, $...$, $2^{N}$, $%
n_{l}=1,0 $ denoting the existence of one or zero particle at the $l$-th
site respectively and $l=1$, $2$, $3$, $...$, $N$. Accordingly the
corresponding state of the Hamiltonian (\ref{tb}), which is transformed from
Eq. (\ref{spin state}), can be expressed explicitly as
\begin{equation}
\left\vert \psi _{f}\right\rangle =\sum_{n}\gamma _{n}\otimes
_{l=1}^{N}[a_{l}^{\dagger }\exp (i\pi \sum_{p=1}^{l-1}\hat{n}%
_{p})]^{n_{l}}\left\vert 0\right\rangle _{f},  \label{fermion state}
\end{equation}%
where $\left\vert 0\right\rangle _{f}$ is the vacuum state, i.e., $%
a_{l}\left\vert 0\right\rangle _{f}=0$.

We now compare the reduced density matrix elements with respect to $%
\left\vert \psi _{s}\right\rangle $ and $\left\vert \psi _{f}\right\rangle $%
. According to Jordan-Wigner transformation (\ref{J-W}), we have
\begin{eqnarray}
\hat{n}_{i}\hat{n}_{j} &=&\frac{1}{4}[1+(\sigma _{i}^{z}+\sigma
_{j}^{z})+\sigma _{i}^{z}\sigma _{j}^{z}], \\
1-\hat{n}_{i}-\hat{n}_{j}+\hat{n}_{i}\hat{n}_{j} &=&\frac{1}{4}[1-(\sigma
_{i}^{z}+\sigma _{j}^{z})+\sigma _{i}^{z}\sigma _{j}^{z}],  \notag \\
a_{i}^{\dagger }a_{j} &=&\frac{1}{4}[\sigma _{i}^{+}\sigma _{j}^{-}\exp
(i\pi \sum_{p=i}^{j-1}\frac{1+\sigma _{p}^{z}}{2})].  \notag
\end{eqnarray}%
Therefore, the reduced density matrix elements for the two systems, which
are defined in Eqs. (\ref{cfs}) and (\ref{cff}), have the following
relations,
\begin{equation}
u^{\pm }=X^{\pm },  \label{u and x}
\end{equation}%
and
\begin{eqnarray}
z &=&\frac{1}{4}\left\langle \psi _{s}\right\vert \sigma _{i}^{+}\sigma
_{j}^{-}\left\vert \psi _{s}\right\rangle ,  \label{difference} \\
Z &=&\left\langle \psi _{f}\right\vert a_{i}^{\dagger }a_{j}\left\vert \psi
_{f}\right\rangle ,  \notag \\
&=&\frac{1}{4}\left\langle \psi _{s}\right\vert \sigma _{i}^{+}\sigma
_{j}^{-}\exp (i\pi \sum_{p=i}^{j-1}\frac{1+\sigma _{p}^{z}}{2})\left\vert
\psi _{s}\right\rangle .  \notag
\end{eqnarray}%
We notice that $z$ is different from $Z$, which is caused by the additional
term $\exp [i\pi \sum_{p=i}^{j-1}(1+\sigma _{p}^{z})/2]$ resulted from the
anti-commutation relation. Thus the magnitudes of C and MC for the two
corresponding states $\left\vert \psi _{s}\right\rangle $ and $\left\vert
\psi _{f}\right\rangle $ may differ from each other. However, no specific
relations between C and MC can be derived from the above equations. In the
following sub section, with the aid of Marshall's sign rule, we will concern
the specific properties of the matrix elements $z$ and $Z$ with respect to
the corresponding ground states of the spin-$1/2$ $XXZ$ chain and the
corresponding spinless fermion model.

\subsection{Specific relations between C and MC for the ground states}

Now we focus on the study of the spin-1/2 $XXZ$\ chain system, of which the
Hamiltonian is in the form of Eq. (\ref{xxz h}). When two specific sites $i$
and $j$ are concerned, the ground state of the spin-1/2 $XXZ$\ chain systems
can be explicitly written as

\begin{eqnarray}
\left\vert \psi _{gs}\right\rangle &=&\left\vert \uparrow \right\rangle
_{i}\left\vert \uparrow \right\rangle _{j}\otimes \left\vert \psi
_{1}\right\rangle +\left\vert \downarrow \right\rangle _{i}\left\vert
\downarrow \right\rangle _{j}\otimes \left\vert \psi _{2}\right\rangle
\label{state of sign} \\
&&+\sum_{k}(x_{k}\left\vert \uparrow \right\rangle _{i}\left\vert \downarrow
\right\rangle _{j}+y_{k}\left\vert \downarrow \right\rangle _{i}\left\vert
\uparrow \right\rangle _{j})\otimes \left\vert \phi _{k}\right\rangle ,
\notag
\end{eqnarray}%
where%
\begin{eqnarray}
\left\vert \psi _{1}\right\rangle &=&\sum_{m}g_{m1}\otimes _{l\neq
i,j}^{N}(S_{l}^{+})^{f(m_{l})}\left\vert 0\right\rangle _{s}, \\
\left\vert \psi _{2}\right\rangle &=&\sum_{n}g_{m2}\otimes _{l\neq
i,j}^{N}(S_{l}^{+})^{f(m_{l})}\left\vert 0\right\rangle _{s},  \notag \\
\left\vert \phi _{k}\right\rangle &=&\otimes _{l\neq
i,j}^{N}(S_{l}^{+})^{f(m_{l})}\left\vert 0\right\rangle _{s},  \notag
\end{eqnarray}%
denoting the basis of the rest $N-2$ sites, and $x_{k}$, $y_{k}$, $g_{m1}$, $%
g_{m2}$ are normalized coefficients. Here we have used the fact, $%
[S^{z},H_{s}^{XXZ}]=0$, which ensures that $\left\vert \psi
_{gs}\right\rangle $ can be written in a single invariant subspace. Then the
correlation function $\left\langle \sigma _{i}^{+}\sigma
_{j}^{-}\right\rangle $ can be expressed as

\begin{equation}
\left\langle \sigma _{i}^{+}\sigma _{j}^{-}\right\rangle =4\left\langle
S_{i}^{+}S_{j}^{-}\right\rangle =4\sum_{k}x_{k}y_{k}.
\end{equation}

Now we will prove an important equation in this paper, i.e., for the spin-$%
1/2$ $XXZ$ chain model (\ref{xxz h}) with arbitrary value of $J_{j}$ and $%
J_{j}^{z},$ it can be verified that for the ground state (\ref{state of sign}%
),%
\begin{equation}
\left\vert \sum_{k}x_{k}y_{k}\right\vert =\sum_{k}\left\vert
x_{k}y_{k}\right\vert .  \label{equation}
\end{equation}%
Before the proof of the above equation, we will first introduce the
Marshall's sign rule \cite{Marshall} for bipartite (i.e., the lattice can be
divided into two sublattices $A$, $B$ such that all nearest neighbors of a
site on one sublattice lie on the other and vice versa) $XXZ$ systems \cite%
{sign rule}. Obviously, the spin-$1/2$ $XXZ$ chain model concerned in the
present paper belongs to a bipartite system. In the following, we will
present the sign rule of a bipartite $XXZ$ model in the situation of
arbitrary $J_{j}^{z}$ and $J_{j}>0$. The sign rule for arbitrary $J_{j}^{z}$
and $J_{j}\leq 0$ is similar to the first situation, and is introduced in
detail in Ref. \cite{sign rule}.

We rewrite the ground state of the bipartite $XXZ$ system as
\begin{equation}
\left\vert \psi _{gs}\right\rangle =\sum_{m}g_{m}\left\vert m\right\rangle
_{s},
\end{equation}%
where $g_{m}$ are normalized coefficients, According to the sign rule \cite%
{sign rule}, in the situation when arbitrary $J_{j}^{z}$ and $J_{j}>0$, the
normalized coefficients $g_{m}$ can be written in the following form

\begin{equation}
g_{m}=(-1)^{\varphi (m)}b_{m},  \label{B situation}
\end{equation}%
where $\{b_{m}\}$ is a positive semi-definite set, and
\begin{equation}
\varphi (m)=\sum_{l\in A}f(m_{l})
\end{equation}%
denotes the number of spin ups in sublattice $A$ for the basis $\left\vert
m\right\rangle _{s}$.

Applying the above sign rule to the ground state (\ref{state of sign}), it
is simply found that the sign of the factor $x_{k}y_{k}$ only depends on the
location of $i$ and $j,$ which is independent of the $k$. If $i$, $j\in A$
or $\notin A$ ($\in B$), we have $x_{k}y_{k}\geq 0$ for any $k$; while if $i$%
, $j$ belong to different sublattices, we have $x_{k}y_{k}\leq 0$ for any $k$%
. Then we have proved the Eq. (\ref{equation}) for arbitrary $J_{j}^{z}$ and
$J_{j}>0$. The situation of arbitrary $J_{j}^{z}$ and $J_{j}\leq 0$ will
give a similar proof. Therefore, the Eq. (\ref{equation}) is valid for
arbitrary $J_{j}^{z}$ and $J_{j}$, which is crucial for the following
discussions.

With the help of sign rules, we now further compare the two measures of
entanglement. Based on the identity (\ref{equation}), we have the following
conclusions for the correlation functions $z$ and $Z$ with respect to the
ground state of spin-1/2 $XXZ$\ model and its counterpart many-particle
ground state in spinless fermion system.

(i) For NN cases, i.e., $j=i+1,$ the\ absolute value of the correlation
function $Z$ is identical to that of the correlation function $z$,

\begin{eqnarray}
\left\vert Z\right\vert &=&\frac{1}{4}\left\vert \left\langle \exp (i\pi
\sum_{p=i}^{i}\frac{\sigma _{p}^{z}+1}{2})\sigma _{i}^{+}\sigma
_{j}^{-}\right\rangle \right\vert \\
&=&\left\vert \exp (i\pi \frac{\sigma _{i}^{z}+1}{2})\sum_{k}x_{k}y_{k}%
\right\vert  \notag \\
&=&\sum_{k}\left\vert x_{k}y_{k}\right\vert =\left\vert z\right\vert ;
\notag
\end{eqnarray}%
\

(ii) For non-NN cases, i.e., $j>i+1,$ the absolute values of the
corresponding correlation functions have the following relation,

\begin{eqnarray}
\left\vert Z\right\vert &=&\frac{1}{4}\left\vert \left\langle \sigma
_{i}^{+}\sigma _{j}^{-}\exp (i\pi \sum_{p=i}^{j-1}\frac{\sigma _{p}^{z}+1}{2}%
)\right\rangle \right\vert \\
&\leq &\sum_{k}\left\vert x_{k}y_{k}\right\vert =\left\vert z\right\vert .
\notag
\end{eqnarray}

Obviously, as we have mentioned in sub section A, the above inequalities are
caused by the anti-commutation relations of fermion systems. Since we have $%
u^{\pm }=X^{\pm }$ as presented in Eq. (\ref{u and x}), the above
inequalities will simply lead to our main observation that there exists a
simple and explicit relation between the concurrences of the two counterpart
ground states, i.e., for NN cases,

\begin{eqnarray}
C_{i,j} &=&2\max \left\{ 0,\left\vert z\right\vert -\sqrt{u^{+}u^{-}}\right\}
\\
&=&(MC)_{i,j}=2\max \left\{ 0,\left\vert Z\right\vert -\sqrt{X^{+}X^{-}}%
\right\} ,  \notag
\end{eqnarray}%
while for non-NN cases, we have $C_{i,j}\geq (MC)_{i,j}$. Therefore, by
combining the two cases, we get the following relation for any types of C
and MC,

\begin{equation}
C_{i,j}\geq (MC)_{i,j}  \label{relation}
\end{equation}%
We emphasize that the above inequality is valid in the ground states of spin-%
$1/2$ $XXZ$ chain models and their spinless fermion counterparts, where the
sign rules hold.

\section{Further comparison in specific models}

The above discussion in Sec. II. C is based on the ground state of a $XXZ$
chain model, of which the counterpart Hamiltonian (\ref{tb}) in spinless
fermion systems corresponds to TB models. The TB models are widely used in
modeling metallic, semiconducting, ionic systems \cite{TB} and recently
optical lattices \cite{optical}, and are of great interest to both the
condensed matter and the quantum information communities. In this section,
we further the above discussion and give an example in the $XY$ chain and TB
models. The $XY$ model is extensively used to describe various quantum spin
systems and exhibit rich quantum phenomena such as the coupling in a
Josephson junction array, quantum order-disorder phase transitions and etc
\cite{JJ}. In this way, the comparison between the ground-state entanglement
of $XY$ models and TB models becomes even more interesting and significant.

As performed in Sec. II. B, the Hamiltonian and the ground state of the $XY$
chain models can be formally transformed into their counterparts in spinless
fermion systems. The Hamiltonian of the $XY$ chain model reads%
\begin{equation}
H_{s}^{XY}=\sum_{j}J_{j}(S_{j}^{+}S_{j+1}^{-}+S_{j}^{-}S_{j+1}^{+}).
\label{xy H}
\end{equation}%
Therefore, the counterpart of the above Hamiltonian after Jordan-Wigner
transformation is given by%
\begin{equation}
H_{f}^{TB}=\sum_{j}J_{j}(a_{j}^{\dagger }a_{j+1}+\text{H.c.}).
\label{free fermion H}
\end{equation}%
It simply corresponds to a spinless fermion hopping model. As we all know if
Eq. (\ref{spin state}) is the ground state of Eq. (\ref{xy H}), then the
ground state of Eq. (\ref{free fermion H}) is given by Eq. (\ref{fermion
state}) in its corresponding particle number sub-space. Therefore, both
ground states of the above two models have practical correspondences, which
will enhance the significance of the specific relations (\ref{relation})
derived in Sec. II. C.

\subsection{An example governed by the sign rule}

We now give an example of the above discussion. Consider a five-site $XY$
chain, of which the Hamiltonian is given as

\begin{eqnarray}
H_{s}^{XY} &=&\sum_{j=1,4}(S_{j}^{+}S_{j+1}^{-}+S_{j}^{-}S_{j+1}^{+})
\label{5 points H} \\
&&+2\sum_{j=2,3}(S_{j}^{+}S_{j+1}^{-}+S_{j}^{-}S_{j+1}^{+})  \notag
\end{eqnarray}%
with the ground state

\begin{eqnarray}
\left\vert \psi _{gs}\right\rangle &=&\frac{1}{6}(-2\left\vert \downarrow
\uparrow \downarrow \uparrow \uparrow \right\rangle _{s}-2\left\vert
\uparrow \uparrow \downarrow \uparrow \downarrow \right\rangle
_{s}+2\left\vert \uparrow \uparrow \downarrow \downarrow \uparrow
\right\rangle _{s} \\
&&+2\left\vert \uparrow \downarrow \downarrow \uparrow \uparrow
\right\rangle _{s}+2\left\vert \uparrow \downarrow \uparrow \uparrow
\downarrow \right\rangle _{s}+2\left\vert \downarrow \uparrow \uparrow
\downarrow \uparrow \right\rangle _{s}  \notag \\
&&+\left\vert \downarrow \downarrow \uparrow \uparrow \uparrow \right\rangle
_{s}+\left\vert \uparrow \uparrow \uparrow \downarrow \downarrow
\right\rangle _{s}-3\left\vert \uparrow \downarrow \uparrow \downarrow
\uparrow \right\rangle _{s}-\left\vert \downarrow \uparrow \uparrow \uparrow
\downarrow \right\rangle _{s}).  \notag
\end{eqnarray}%
We calculate the concurrence between site $1$ and $3$, which results in $%
u^{+}=7/18,$ $u^{-}=1/9,$ $z=2/9,$ and therefore,

\begin{eqnarray}
C_{1,3} &=&2\max \left\{ 0,\left\vert z\right\vert -\sqrt{u^{+}u^{-}}\right\}
\\
&=&\frac{1}{9}(4-\sqrt{14}).  \notag
\end{eqnarray}%
On the other hand, we\ transform the ground state of $XY$ chain model (\ref%
{5 points H}) into its spinless fermion counterpart, which is accordingly
the ground state of the counterpart Hamiltonian of Eq. (\ref{5 points H}).
As a result, the correlation functions are given as $X^{+}=7/18$, $X^{-}=1/9$%
, $Z=1/9$ for the counterpart ground state, thus we have

\begin{eqnarray}
(MC)_{1,3} &=&2\max \left\{ 0,\left\vert Z\right\vert -\sqrt{X^{+}X^{-}}%
\right\} \\
&=&0.  \notag
\end{eqnarray}%
Consequently, we get that $C_{1,3}>(MC)_{1,3}$, which is in agreement with
Eq. (\ref{relation}). Here the difference between $z$ and $Z$ is apparently
resulted from the different commutation relations of the two models.

We remark that our theoretical proof in Sec. II is valid in the above
example, which relies on the Marshall's sign rule. However, we are surprise
to notice that the inequality (\ref{relation}) also holds for excited states
and even for eigen states of many other models. These situations are beyond
our theoretical proof and we will give some examples in the following
subsection. Thus the relation (\ref{relation}) may provide us a deeper and
more general insight into the differences between spin-$1/2$ and fermion
systems.

\subsection{Examples where the sign rule does not apply}

We consider the situations where the sign rule is violated. Here we take a
TB model with uniform hopping integral as an example, of which the
Hamiltonian reads
\begin{equation}
H_{f}^{TB}=\sum_{l}^{N-1}(a_{l}^{\dagger }a_{l+1}+\text{H.c.}).
\end{equation}%
The single-particle eigenstates are $\left\vert k\right\rangle =\sqrt{2/(N+1)%
}\sum_{l}^{N}\sin (kl)a_{l}^{\dagger }\left\vert 0\right\rangle $, where $%
k=n\pi /(N+1)$ and $n=1,2,...,N$, with the corresponding spectrum $\epsilon
_{k}=2\cos k$. According to the above analysis, an arbitrary two-particle
state can be written as
\begin{equation}
\left\vert k,k^{\prime }\right\rangle =\sum_{l<l^{\prime }}^{N}D(k,k^{\prime
},l,l^{\prime })a_{l}^{\dagger }a_{l^{\prime }}^{\dagger }\left\vert
0\right\rangle ,
\end{equation}%
where

\begin{equation}
D(k,k^{\prime },l,l^{\prime })=\frac{2}{N+1}\det \left\vert
\begin{array}{cc}
\sin (kl) & \sin (kl^{\prime }) \\
\sin (k^{\prime }l) & \sin (k^{\prime }l^{\prime })%
\end{array}%
\right\vert .
\end{equation}%
Assuming $i<j,$ we have
\begin{equation}
Z=\sum_{l}D(k,k^{\prime },j,l)D(k,k^{\prime },i,l).  \label{Z1}
\end{equation}

According to the commutation relations, the corresponding correlation
function in the $XY$ spin model can be simply written as
\begin{eqnarray}
z &=&\frac{1}{4}\left\langle \sigma _{i}^{+}\sigma _{j}^{-}\right\rangle
\label{Z2} \\
&=&(\sum_{l<i}-\sum_{i<l<j}+\sum_{l>j})D(k,k^{\prime },j,l)D(k,k^{\prime
},i,l).  \notag
\end{eqnarray}

Now we consider the sign of three terms in $z$ corresponding to the $XY$
spin model. We take $k=\pi /(N+1)$ and $k^{\prime }=2\pi /(N+1)$ as an
example. Obviously, the state $\left\vert k,k^{\prime }\right\rangle $ is an
excited state, which does not obey the Marshall's sign rule. A
straightforward calculation gives

\begin{equation}
D(k,k^{\prime },j,l)D(k,k^{\prime },i,l)>0,
\end{equation}%
for any $l\neq i,j$. Therefore, from Eqs. (\ref{Z1}) and (\ref{Z2}), we have
\begin{equation}
\left\vert z\right\vert >\left\vert Z\right\vert \Longrightarrow C_{i,j}\geq
(MC)_{i,j},
\end{equation}%
which shows that our conclusion still holds even the Marshall's sign rule is
violated.

Now we consider another excited state with $k=\pi /(N+1)$ and $k^{\prime
}=N\pi /(N+1)$, which is also beyond the Marshall's sign rule. For even $j$
or $i$, we have
\begin{eqnarray}
&&D(k,k^{\prime },j,l)D(k,k^{\prime },i,l) \\
&=&D(k,\pi -k,j,l)D(k,\pi -k,i,l)  \notag \\
&=&0.  \notag
\end{eqnarray}%
which leads to
\begin{equation}
\left\vert z\right\vert =\left\vert Z\right\vert =0\Longrightarrow
C_{i,j}=(MC)_{i,j}
\end{equation}

Although both of the above two examples are excited states, which violate
the Marshall's sign rule, they are still in agreement with the conclusion of
Eq. (\ref{relation}). Thus it will be very interesting to give an even more
general proof to extend our conclusion for the states that beyond the
Marshall's sign rule.

\section{Summary and discussion}

As a conclusion we have compared the concepts of concurrence for spin-$1/2$
systems with mode concurrence for spinless fermion systems explicitly. By
employing the Jordan-Wigner transformation and the Marshall's sign rule, we
come to our main observations in spin-$1/2$ $XXZ$ and spinless fermion chain
systems that: concurrence and mode concurrence are different from each other
for general corresponding states of the two systems and further there exist
specific relations between the ground-state concurrence of a spin-$1/2$ $XXZ$
chain model and the mode concurrence of its counterpart ground state in a
many-particle spinless fermion model. (i) The nearest neighbor ground-state
concurrence of spin-$1/2$ $XXZ$ chain models is identical to the nearest
neighbor ground-state MC of many-particle spinless fermion systems, i.e., $%
C_{i,i+1}=(MC)_{i,i+1}$. (ii) For other types of entanglement, concurrences
are no less than moce concurrences for any given corresponding ground
states, i.e., $C_{i,j}\geq (MC)_{i,j}$ for $j>i+1$. An example in $XY$ spin
chain model and spinless fermion hopping model with practical significance
is given to illustrate the simple relation derived from the comparison
between concurrence and mode concurrence. The differences between the
ground-state entanglement of the two models are closely related to the
fundamental features (commutation relations) of the two models and also may
indicate some new aspects of the intrinsic distinctions between spin-$1/2$ $%
XXZ$ and spinless fermion chain systems. We have also shown some other
states and examples that are beyond our analytical proof and we observe that
all these results are in agreement with the relation (\ref{relation}). Thus
it will be very interesting to further investigate and to generalize the
relations between the measures of entanglement in spin-$1/2$, fermion and
even boson systems, as well as the intrinsic properties of the these systems.

We gratefully acknowledge the valuable discussion with Professor Chang-Pu
Sun. This work is supported by the CNSF (Grant No. 10474104), the National
Fundamental Research Program of China (No. 2001CB309310).

\end{document}